\documentclass[sigconf,9pt]{acmart}

\usepackage{booktabs,multicol,multirow}
\usepackage{algorithm}
\usepackage{fancyhdr}

\copyrightyear{2026}
\acmYear{2026}
\setcopyright{cc}
\setcctype{by-nc-nd}
\acmConference[GLSVLSI '26]{Great Lakes Symposium on VLSI 2026}{June 22--24, 2026}{Canandaigua, NY, USA}
\acmBooktitle{Great Lakes Symposium on VLSI 2026 (GLSVLSI '26), June 22--24, 2026, Canandaigua, NY, USA}
\acmDOI{10.1145/3787109.3815215}
\acmISBN{979-8-4007-2431-2/2026/06}

\begin{document}

\title[ObfAx: Obfuscation and IP Piracy Detection in Approximate Circuits]{ObfAx: Obfuscation and IP Piracy Detection in Approximate Circuits}

\author{Lukas Sekanina}
\orcid{0000-0002-2693-9011}
\affiliation{%
\institution{Brno University of Technology} %
\department{Faculty of Information Technology} %
 \city{Brno}
 \country{Czech Republic}
}
\email{sekanina@fit.vut.cz}

\author{Vojtech Mrazek}
\orcid{0000-0002-9399-9313}
\affiliation{%
\institution{Brno University of Technology} %
 \department{Faculty of Information Technology} %
 \city{Brno}
 \country{Czech Republic}
}
\email{mrazek@fit.vutbr.cz}
\begin{abstract}
Approximate circuits often achieve exceptional trade-offs between computational accuracy and hardware efficiency, making them attractive for deployment as reusable Intellectual Property (IP) cores. However, safeguarding such circuits against piracy is critical for enabling sustainable commercialization of approximate computing. This work addresses the emerging challenge of IP protection and piracy detection in the context of approximate hardware. We introduce a novel adversarial threat model, approximate obfuscation, in which an attacker not only conceals the design through structural obfuscation but also introduces functional modifications to ensure that the resulting circuit exhibits nearly identical error characteristics and hardware metrics as the original IP. 
To counter this threat, we propose an automated framework that extracts and compares statistical error profiles of protected IP cores and suspicious circuits, enabling systematic detection of potential IP theft. Through extensive experiments on a diverse set of approximate multipliers, we analyze the resilience of different approximate multipliers against approximate obfuscation. Our results provide new insights into the interplay between obfuscation, approximation, and IP protection.
\end{abstract}


\maketitle

\pagestyle{fancy}
\fancyhf{}
\fancyhead[C]{\small\itshape To appear at the Great Lakes Symposium on VLSI 2026 (GLSVLSI '26)}
\renewcommand{\headrulewidth}{0.4pt}
\fancyfoot[C]{\thepage}
\fancypagestyle{firstpagestyle}{%
  \fancyhf{}
  \fancyhead[C]{\small\itshape To appear at the Great Lakes Symposium on VLSI 2026 (GLSVLSI '26)}
  \renewcommand{\headrulewidth}{0.4pt}
  \fancyfoot[C]{\thepage}
}

\section{Introduction}

Approximate circuits are digital designs intentionally modified or simplified to reduce hardware complexity, power consumption, or delay at the expense of losing exact functional correctness~\cite{ACsurvey:ACM:2020}. 
Designing a high-quality approximate circuit is a non-trivial and resource-intensive process.
A designer must explore a wide range of possible approximations to identify implementations that meet tight, application-specific constraints on error metrics while simultaneously optimizing hardware parameters such as area, power, and delay.
This process typically involves iterative simulation, formal verification, and synthesis runs, often guided by computational intelligence methods or manual expert tuning---sometimes requiring days or weeks of engineering effort~\cite{ceska:iccad17,Mrazek:evoapproxlib:2017}.
The resulting circuits thus represent valuable Intellectual Property (IP) assets whose unauthorized reuse should be reliably detectable.
Many approximate circuits are currently available at the level of netlists or Verilog/VHDL source files in open source libraries~\cite{Shafique:dac16, Mrazek:evoapproxlib:2017, SMApproxlib:dac18, Scarabottolo:pieee:2020}. 
They often achieve exceptional trade-offs between accuracy and hardware efficiency, making them attractive for deployment as reusable IP  cores. 
The available literature discusses various security issues of approximate circuits~\cite{Regazzoni:iccad2018, Liu:2020, MAAS:trojan:axc:2023}; however, the problem of IP protection and IP piracy detection has not been adequately addressed. 

A central challenge addressed in this paper is how to reliably detect the theft or misuse of approximate circuits. Let $\alpha$ be an approximate circuit that is offered as a protected IP core. If the task were to detect whether a suspicious circuit $\beta$ is an exact copy or an obfuscated version of a protected approximate circuit $\alpha$, standard techniques for IP theft detection~\cite{Rostami:2014, GNN4IP:2021} can be applied. 
However, the errors in behavior, which are an inherent manifestation of approximate circuits, make IP detection significantly different from the scope of exact circuits and thus difficult. We propose to consider a circuit $\beta$ created by a \textit{short series of small modifications}  of $\alpha$ is an \textbf{IP theft}. By a small modification, we mean a modification (such as gate deletion, introduction, or reconnection) introduced into the circuit structure to mask the implementation of original circuit $\alpha$. 
We thus introduce a novel adversarial threat model---\textbf{approximate obfuscation}---in which an attacker not only conceals the design through functionality-preserving obfuscation but also introduces functional modifications. The attacker's aim is to ensure that the infringing circuit exhibits very close error characteristics and hardware metrics to the original IP because they are crucial for its deployment.

\begin{figure}[b]
   \centering
   \vspace{-1.5em}
   \includegraphics[width=0.9\linewidth]{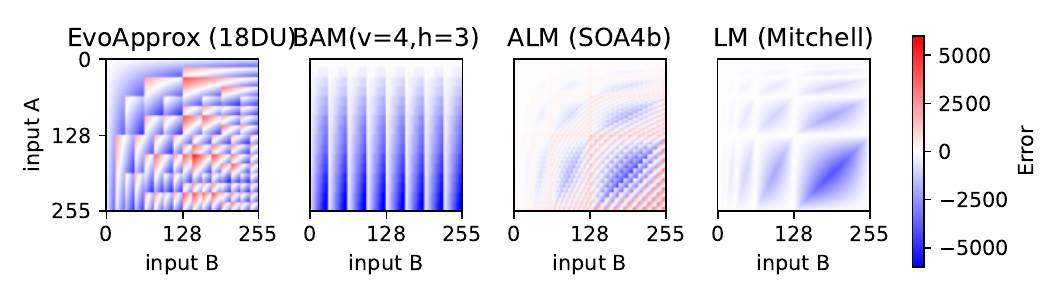}
   \vspace{-2em}
   
   \caption{Family-specific error profiles of selected approximate 8-bit multipliers with similar worst-case error.}
   \label{fig:map:errdist}
\end{figure}
The key intuition behind the proposed detection is illustrated in Fig.~\ref{fig:map:errdist}: although multipliers from different families~\cite{JiangLL019:bookch,Jiang:axc:surv:2020} may share nearly identical worst-case error, their error distributions differ markedly in structure.
Logarithm-based multipliers (ALM, Mitchell) produce smooth globally structured patterns, regular array-based designs (BAM) yield characteristic striped artifacts, and evolutionary designs (EvoApproxLib) display irregular noise-like distributions.
These family-specific signatures persist even after structural obfuscation (see Section~\ref{sec:res:obf}), forming the fingerprint that can be utilized for IP theft detection.

In this paper, we propose and evaluate new techniques that can effectively support the detection of IP theft in the context of approximate arithmetic circuits.
The decision whether an approximate circuit $\beta$ was created from an approximate circuit $\alpha$ can be based on a comparison of behavior, netlist, or both behavior and netlist of these circuits.  
We assume that only circuit behavior (no netlist) is available for analysis. By showing that the behavior of $\beta$ and $\alpha$ is very similar (similarity analysis will be conducted in Section~\ref{sec:sim:analyz}), one can quickly identify a suspicious circuit and send it for a subsequent detailed structural analysis if a netlist is available. 

To detect that circuit $\beta$ was created from $\alpha$ by approximate obfuscation, we compute error heat maps and other error metrics for both circuits and use them as inputs to a binary classifier.
We compare three approaches to creating the classifier, including standard statistical machine learning classifiers and a specialized neural network called Siamese neural network~\cite{SiameseDNN}. To show that the proposed strategy provides reliable results, we use a large collection of existing and obfuscated approximate multipliers to train and test the classifiers. To construct this set automatically, i.e., to efficiently model the behavior of a thief, we employ 
Cartesian genetic programming (CGP)~\cite{miller:cgp:book}. The task is to find good approximate obfuscations, i.e., structurally modified circuits showing a very similar error and hardware properties compared to an original approximate circuit $\alpha$.  
In summary, the paper makes the following contributions:
\begin{itemize}\vspace{-0.3em}
  \setlength{\leftskip}{-2em}
\item We introduce a new adversarial threat model, approximate obfuscation, and a CGP-based algorithm that effectively generates obfuscated approximate circuits showing similar error and hardware properties to protected circuits.
\item We analyze several types of approximate multipliers and show that some of them, particularly circuits from the EvoApproxLib library~\cite{Mrazek:evoapproxlib:2017}, are significantly more resistant to approximate obfuscation. 
\item We develop a data set containing 1,775 approximate multipliers generated using the proposed obfuscation method from 97 existing approximate 8-bit multipliers. 
\item Using a suitable classifier, we can effectively detect obfuscated approximate 8-bit multipliers with accuracy greater than 97\% in our data set. 
\end{itemize}\vspace{-1em}

\section{Related Work}

\subsection{Approximate Circuits}

Within the scope of approximate computing~\cite{ACsurvey:ACM:2020}, we restrict ourselves to 
approximate integer multipliers. We consider functional approximation techniques only because they are the most frequently utilized in this area.
A recent survey~\cite{Wu:axmult:acm:surv:2024} distinguishes three main levels when developing approximate multipliers. 
At the \emph{algorithm level}, the exact standard multiplication is approximated by another algorithm that produces inexact but sufficiently good results (e.g., a logarithm-based multiplier (ALM)~\cite{ALM}).
At the \emph{architecture level}, a suitable exact implementation is simplified.
A typical example is a broken array multiplier (BAM) in which the partial product array is divided into four groups by horizontal ($h$) slicing and vertical ($v$) slicing~\cite{Mahdiani:TCSI2009}. A particular setting of $v$ and $h$ determines which part of the array is ignored to save chip area. At the \emph{circuit level}, gate- and transistor-level approximation techniques such as circuit rewriting and various machine learning-based (ML) methods are applied~\cite{Mrazek:evoapproxlib:2017, WitschenMAP19, Yi:GPTAC:2024} to create new circuit structures.

The design of approximate circuits is driven by error metrics, e.g.,
worst-case error (WCE), mean absolute error (MAE), and error probability (EP).
One of the automated circuit design methods whose roots are in ML is CGP~\cite{miller:cgp:book}.
It has been used to design approximate arithmetic circuits for EvoApproxLib~\cite{Mrazek:evoapproxlib:2017} and other circuits~\cite{Wille:cgp:2024}. %
According to~\cite{Mrazek:evoapproxlib:2017, ceska:iccad17}, an original (exact) gate-level circuit is selected as a seed, which is modified using some perturbations (mutations) to reach, after some steps, a circuit with the desired properties.  
For example, Ceska et al.~\cite{ceska:iccad17} minimizes the relative circuit area (which highly correlates with power consumption) under a user-prescribed constraint on the WCE. Another search strategy for approximate circuit design was used in~\cite{WitschenMAP19}.

\subsection{Security in the Context of Approx. Circuits}

The papers~\cite{Liu:2020, Regazzoni:iccad2018} provide an overview of relevant security issues, emphasizing the fact that approximate computing systems can exhibit uncertainty and unpredictable behavior, which is hard to distinguish from intentional undesired manipulation of the system by an attacker. For example: (i) The approximate computation and storage modules could be manipulated to establish a covert channel through which sensitive information is leaked. (ii) An advanced circuit design method can integrate hardware Trojans into approximate hardware accelerators. The infected Trojans are difficult to identify using conventional measurement techniques~\cite{MAAS:trojan:axc:2023}. 
(iii) A recent study~\cite{Yellu:2023} addresses security threats arising from the unprotected boundary between approximate and precise modules.
On the other hand, approximate computing can help improve security because it can provide efficient implementations of various security mechanisms, enable new information hiding schemes, or increase the performance of ML-based attack identification methods~\cite{Liu:2020}. 

\subsection{IP Piracy}

By Intellectual Property (IP) piracy, we mean a situation where the IP is stolen (extracted from a larger design or IP library in a foundry, online repository, final hardware product, etc.) or used without the knowledge and consent of the owner~\cite{Rostami:2014}. In the case of conventional (exact) circuits, IP protection has become a cornerstone of the modern semiconductor supply chain. However, no attention has been paid to protecting approximate circuits from illegal use.

Common techniques developed to thwart IP piracy include watermarking, fingerprinting, hardware metering, logic locking, obfuscation, and camouflaging~\cite{Sandip:2017,Rostami:2014,GNN4IP:2021}. These techniques introduce additional hardware overhead and cannot guarantee security against modern attacks (such as Boolean satisfiability attacks~\cite{antiSAT:attack:2019}). 

Recent IP piracy detection approaches, such as GNN4IP~\cite{GNN4IP:2021}, contrasted with insertion signatures and other mechanisms that add hardware overhead, create circuit models, and assess the similarity between these models using advanced deep neural networks. 
However, this approach is hard to apply in the context of approximate computing, where the original and obfuscated circuits are not functionally equivalent. Moreover, the circuit netlist must be available to the method.

\section{Proposed Method}

\begin{figure}[b]
    \centering\vspace{-1em}
    \includegraphics[width=0.9\columnwidth]{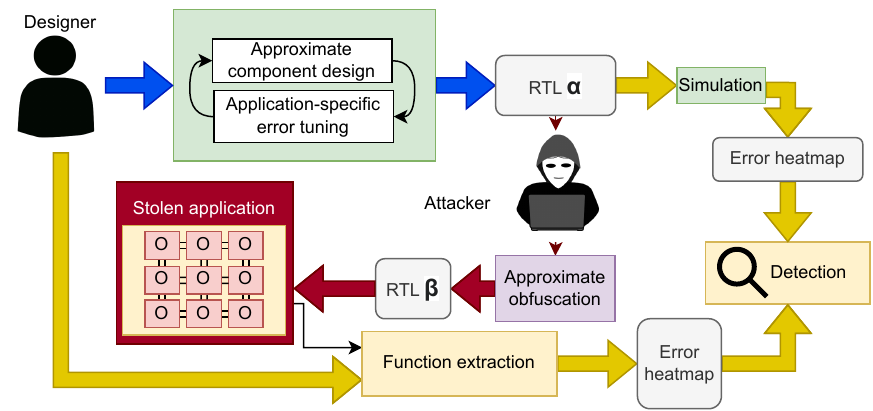}
    \vspace{-1.5em}
    \caption{Threat model. The Designer creates an approximate circuit $\alpha$ for a specific application. The Attacker gets its RTL and uses it with some modifications (circuit $\beta$) in another application. To detect this suspicious circuit, its functionality is extracted and compared with circuit $\alpha$.}
    \label{fig:threat}
\end{figure}

We propose extending the definition of IP piracy to include counterfeit \emph{obfuscated approximate circuits} whose behavior deviates slightly from that of the original approximate circuits. 

\subsection{Conceptual Framework and Its Assumptions}

\textbf{An attacker}: 
Instead of using a common functionality-preserving obfuscation method, the attacker can employ an \emph{approximate obfuscation}, that is, a method that hides the functionality and implementation of $\alpha$ by inserting additional gates into it or reconnecting or modifying existing gates in such a way that $\alpha$ and $\beta$ are not functionally equivalent, but the error is kept acceptable for the target application. Because $\alpha$ and $\beta$ are structurally different and not functionally equivalent, it is potentially difficult for the owner to prove that $\alpha$ or its clones are used without permission.

We assume that an attacker has access to and can modify a gate-level netlist of $\alpha$ in such a way that some gates are removed, added, or reconnected to create $\beta$. However, the number of particular modifications is limited because the attacker does not want to build a new implementation expensively. Their aim is to reuse $\alpha$ as much as possible, sufficiently hide the implementation of $\alpha$, and keep the error and hardware metrics under acceptable bounds.

\textbf{IP theft detection}: The owner of $\alpha$ wants to know whether a suspicious circuit is an exact copy of $\alpha$, an approximately obfuscated version of $\alpha$, or a different circuit. 
In this paper, we assume that the circuit theft detection is based on analysis of the circuit's \textit{functional behavior}, i.e., the suspicious netlist is not available. The motivation for this approach is twofold: (1) Contrasted to exact circuits, it makes sense to perform such an analysis for approximate circuits. It is assumed that the error profiles of $\alpha$ and its obfuscated version $\beta_k$ created using $k$ elementary modifications of $\alpha$ will be similar (under a suitable metric). If so, $\beta_k$ will be marked as possible theft, and a detailed analysis will be performed, potentially considering also the netlist. (2) Obtaining circuit behavior (input-response vectors) can be much easier than obtaining the netlist in some applications. For example, by a suitable configuration of a configurable DNN accelerator, one could set arbitrary inputs for a particular approximate multiplier and read its outputs to reconstruct its truth table at least partially. The threat model is visualized in Fig.~\ref{fig:threat}.

\subsection{Approximate Obfuscation}
\label{sec:cgp:obf}

\begin{figure}[b]
   \centering
   \vspace{-1.5em}
   \includegraphics[width=0.9\linewidth]{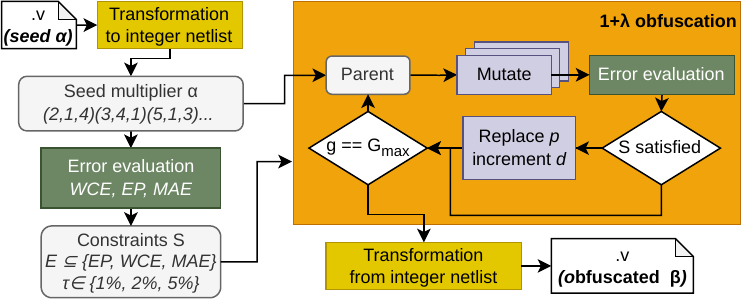}
   \vspace{-1em}
   \caption{Overall schema of the proposed approximate obfuscation algorithm}
   \label{fig:overall}
\end{figure}

Common obfuscation methods preserve functionality, but hide the implementation by adding redundant gates or replacing selected subcircuits with functionally equivalent but structurally different circuits. In our scenario, the attacker can also change the circuit functionality but the obfuscated circuit must show a very similar error and hardware properties.
It is assumed that the attacker obtained a gate-level representation of an approximate circuit to be obfuscated. 
To effectively perform the approximate obfuscation, we propose an automated approach inspired by CGP illustrated in Fig.~\ref{fig:overall}. 
In any case, the space of approximate circuits which can be obtained by applying a series of modifications to $\alpha$ is explored. If properties of a candidate modification remain in the desired range, the candidate circuit is accepted and serves as a new parent for the next iteration.

\textit{Circuit representation:} An original $n_i$-input/$n_o$-output approximate circuit $\alpha$ is modeled using a simplified netlist consisting of two-input gates. This netlist is converted to the CGP chromosome whose size is $3z+n_o$ genes (integers), where $z$ is the number of gates in $\alpha$. For every gate, encoding the destinations for its two inputs requires two genes; one gene is devoted to the gate function. Primary outputs are determined using the $n_o$ genes.

\textit{Mutation:} The chromosome is modified at $h$ randomly determined positions. For each position, a new but valid gene value is generated at random. Note that a mutation can have an effect on neither the circuit function nor the structure. 

\textit{Search strategy:} The so-called 1+$\lambda$ search strategy is adopted~\cite{miller:cgp:book}. The seed ($\alpha$, the so-called parent) is used to generate $\lambda$ offspring chromosomes by mutation. All offspring are evaluated using the objective function. The following steps are repeated ${G}_{max}$ times: (i) the best-scored circuit (i.e., the  new parent) is selected from the previous generation; (ii) $\lambda$ offspring circuits are created from the new parent using mutation; (iii) all offspring circuits are evaluated. 

\textit{Objective function:} 
The objective is to maximize the number of modifications (that is, an edit distance measure at the level of circuit encoding), after which the obfuscated circuit ($\beta$) still exhibits the key properties in the requested range. A candidate circuit $a$ is evaluated using the objective function $F$:
\begin{equation}\vspace{-2mm}
F(a) = 
\begin{cases}
d(a, \alpha) & \text{if } S \text{ is satisfied\ } \\
0 & \text{otherwise,} \\
\end{cases}
\label{eq:fitness}
\end{equation}
where $d(a, \alpha)$ is the edit distance between $a$ and $\alpha$ at the encoding level and $S$ is a set of constraints on the error which must be satisfied:
$E_i(a) \leq E_i(\alpha)\cdot (1 + \tau)$.
Every error metric $E_i$, $i = 1, \dots, s$, (for example, WCE, MAE, or EP as defined in~\cite{vasicek:access2019}) is calculated and compared to the error threshold depending on $\tau$ -- an acceptable error deviation.
Because the maximum number of gates considered by CGP is defined by the number of gates in the seed ($z(\alpha)$), the number of gates in the obfuscated circuit $z(a)$ is always equal to or less than $z(\alpha)$.

The search algorithm generates a sequence of circuit modifications $\alpha, \dots, a, \dots, \beta$ starting with the seed $\alpha$ and finishing with an obfuscated circuit $\beta$. Formally, $\alpha \vdash_S^* \beta$, where a single modification is modeled using a unary relation $\vdash_S$. 
The relation $\vdash_S$ is a subset of $\mathbb{A} \times \mathbb{A}$, where $\mathbb{A}$ is a set of $n_i$-input/$n_o$-output approximate circuits and $S$ is a set of constraints. For $a_1, a_2 \in \mathbb{A}$, $a_1 \vdash_S a_2$ holds iff $S(a_1) \wedge S(a_2) \wedge d(a_1, a_2) = 1$.
The proposed method is executed several times with different settings of parameters and constraints to find the best approximate obfuscation(s) of $\alpha$.

\subsection{Detection of Approximate Obfuscation}

\begin{figure}[b]
    \centering
    \vspace{-1em}
  \includegraphics[width=0.95\linewidth]{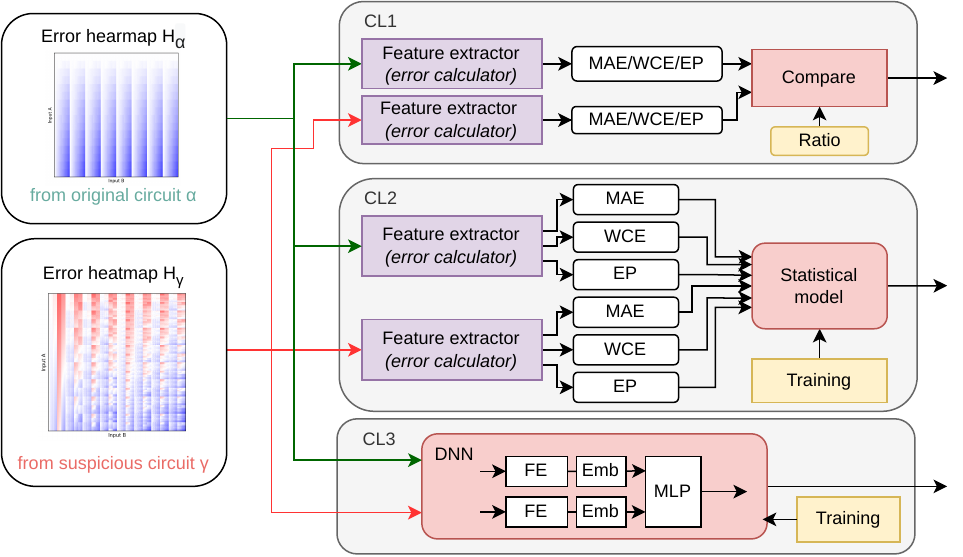}
  \vspace*{-1em}
    \caption{Three approaches to the IP theft detection based on the heatmaps $H_\alpha$ and $H_\gamma$ of original and suspicious circuit.}
    \label{fig:classifier}
\end{figure}

The proposed IP theft detection method assumes that one can obtain or estimate an error heat map $H_\gamma$ for the suspicious approximate circuit $\gamma$. 
For example, for an approximate multiplier and any input operands $i$ and $j$, a two-dimensional error heat map is computed as $H_\gamma(i,j) = \gamma(i, j) - i*j$. An analogous heat map $H_\alpha$ is constructed for $\alpha$. Owing to the method’s partial error resilience, it remains robust even when a substantial fraction of input–output pairs is unavailable. The effect of missing entries on detection accuracy is analyzed in Section~\ref{sec:class}.

The detection method examines a pair of error heat maps  $H_\alpha$ and $H_\gamma$. The IP theft detection problem is formulated as a binary classification problem that decides whether $H_\gamma$ is generated from $H_\alpha$ (Class-Y) or not (Class-N). Fig.~\ref{fig:classifier} illustrates three approaches we propose to the classifier design:
\begin{itemize}

  \setlength{\leftskip}{-2em}
\item Classifier(s) CL1 computes a ratio between MAEs (or WCEs or EPs) obtained from $H_\alpha$ and $H_\gamma$, and applies a simple thresholding to determine the class.  
\item Classifier CL2 computes the standard error metrics (MAE, WCE, EP) from $H_\alpha$ and $H_\gamma$ and uses them as six features to a common statistical machine learning model, which is trained to perform the classification.
\item Classifier CL3 accepts the error heat maps and uses them directly as two input images to a deep neural network (DNN), in particular, a Siamese neural network~\cite{SiameseDNN}. %
\end{itemize}

The output of each classifier is interpreted using standard measures: accuracy, sensitivity, and specificity. Composition of training and test data sets will be discussed in Section~\ref{sec:class}. The fact that approximate multipliers are not necessarily commutative must be reflected in these data sets and during the evaluation.

\section{Results}

\subsection{Generating Approximate Obfuscations}\label{sec:res:obf}

\textit{Data set:} 
We use 75 human-created approximate 8-bit multipliers according to the implementations described in~\cite{JiangLL019:bookch,Wu:axmult:acm:surv:2024} (the number of different multiplier configurations that lead to different error-area tradeoffs is given in parentheses): 
error-tolerant multiplier (ETM, 3 configurations), 
approximate error accumulation scheme 1 (AM1, 15) and scheme 2 (AM2, 15), 
under-designed multiplier (UDM, 1),   
static segment method (SSM, 3),
broken array multiplier (BAM, 21),
approximate logarithm-based multiplier (ALM, 15),
inaccurate multiplier (ICM, 1), and Mitchell approximate multiplier (Mitchell, 1). In addition, 22 approximate 8-bit multipliers are taken from EvoApproxLib~\cite{Mrazek:evoapproxlib:2017}. 

\textit{Setting of the search method:} The design space search algorithm operates with the parameters that we have determined based on several test runs and the setting used in~\cite{Mrazek:evoapproxlib:2017, ceska:iccad17, Wille:cgp:2024}: $n_i = 2*8=16$, $n_o=16$, $\lambda = 1, h = 1, {G}_{max} = 15k$ (corresponding to 3 min per run on average on a single core of Intel Xeon 5670 at 2.93 GHz), i.e., the cost of attack is relatively low.
Three levels of $\tau$ are considered for WCE, MAE, and EP; $\tau = \{0.01, 0.02, 0.05\}$, i.e. 1\%, 2,\%, and 5\%.

\begin{figure}[b]
    \centering
    \vspace{-2.5em}
    \includegraphics[width=\linewidth]{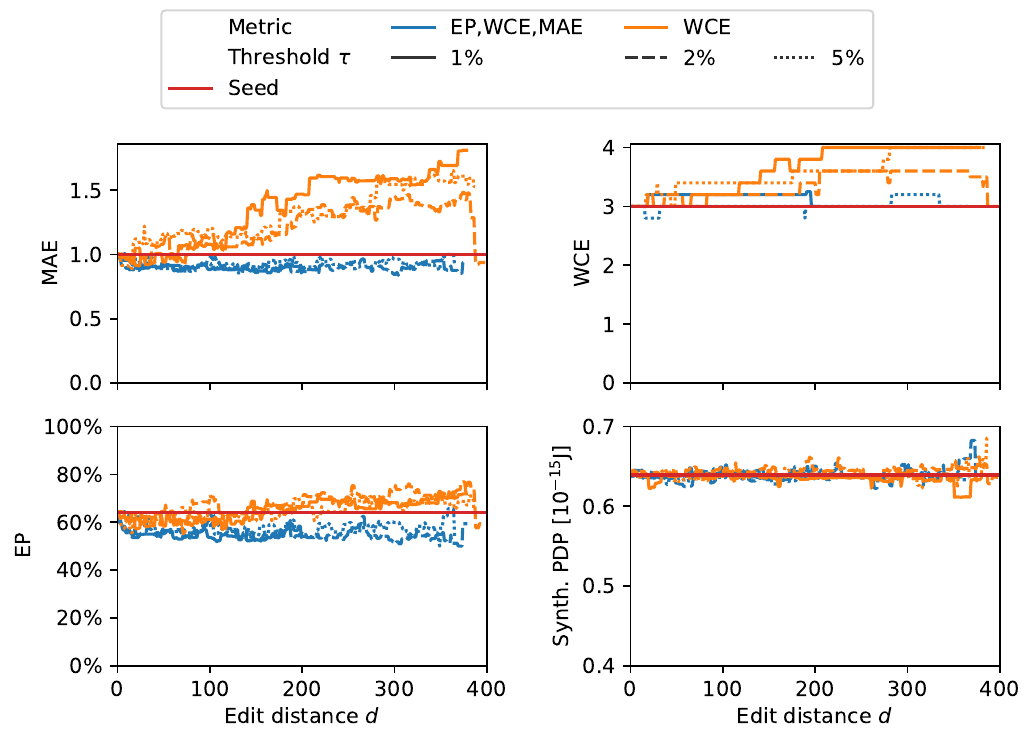}%
    \vspace*{-1.5em}%
        \caption{Progress of error metrics and PDP (averaged for 5 runs for each configuration) during obfuscation of EvoApprox mul8u\_2P7 multiplier constrained for WCE only (orange) and all error metrics (WCE, EP, and MAE) (blue).} 
    \label{fig:evolution}
\end{figure}

\textit{Generating obfuscated multipliers:} For each of the 75 human-designed approximate 8-bit multipliers in our data set and three error thresholds $\tau$, we conducted 5 independent runs of CGP, resulting in 1,125 obfuscated multipliers. Additionally, initializing the search with 22 approximate multipliers from EvoApproxLib under the same conditions produced 650 additional unique obfuscated multipliers. In total, we generated 1,775 obfuscated multipliers.

\begin{figure}[t]
    \centering
    \includegraphics[width=\linewidth]{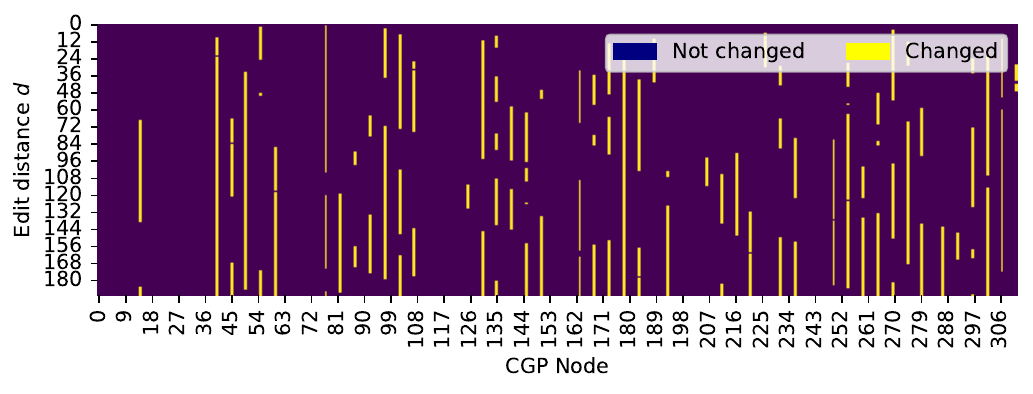}
    \vspace*{-3.0em}
        \caption{Development of changes in the CGP nodes of the mul8u\_2P7 multiplier (consisting of 312 gates) during a single run of the obfuscation algorithm.}
    \vspace{-2em}
    \label{fig:changes}
\end{figure}

\textit{The role of error constraints:} Our initial experiment evaluates the ability of the proposed obfuscation algorithm to generate functional modifications while adhering to specified error constraints. Fig.~\ref{fig:evolution} illustrates the algorithm’s progression during the obfuscation of the mul8u\_2P7 multiplier, which is optimized for WCE and area in EvoApproxLib. When only the WCE is constrained (orange curves), the remaining error metrics exceed their respective tolerance bounds for all considered values of $\tau$ (dashed lines). In contrast, when all error metrics are simultaneously constrained, the errors remain within the required limits (blue curves). Fig.~\ref{fig:evolution} also shows that the Power Delay Product (PDP) we obtained after synthesis of obfuscated circuits using Synopsys design compiler to a 45 nm technology remains within the desired range.

Figure~\ref{fig:changes} presents a detailed analysis of a single obfuscation run for the mul8u\_2P7 multiplier with $\tau=1\%$ and all metrics constrained. The visualization indicates the edit steps at which each CGP node (i.e., two-input gate) was modified during the obfuscation process. Only modifications that affected the error are shown. In some cases, earlier modifications were later reverted. %

\begin{figure}[ht]
    \centering
    \includegraphics[width=\linewidth]{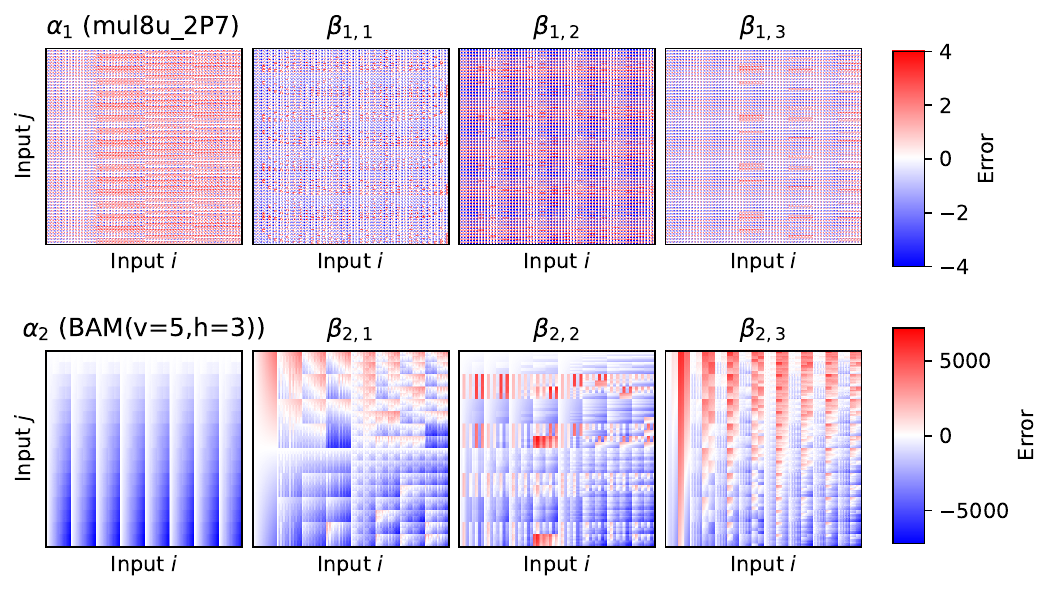}%
    \vspace*{-1.5em}%
    \caption{Error heatmaps of two approximate multipliers used as seeds and their three obfuscated versions: mul8u\_2P7 (top) and BAM($v=5,h=3$) (bottom).}
    \label{fig:example}
\end{figure}

\textit{Examples of results:} Figure~\ref{fig:example} illustrates error profiles for two distinct approximate multipliers that we used as seeds ($\alpha$ circuits): mul8u\_2P7 exhibiting near-random noise behavior and a structured BAM multiplier ($v=5, h=3$). The error profiles of three randomly selected obfuscated variants with comparable error properties are shown for each seed. For BAM multipliers, the regular structural patterns remain preserved while the error range shifts to include positive errors (shown in red). In contrast, no discernible patterns are visible in the error profiles of obfuscated multipliers created from mul8u\_2P7 (a circuit with an inherently irregular structure).

\begin{figure}
    \centering
    \includegraphics[width=0.9\columnwidth]{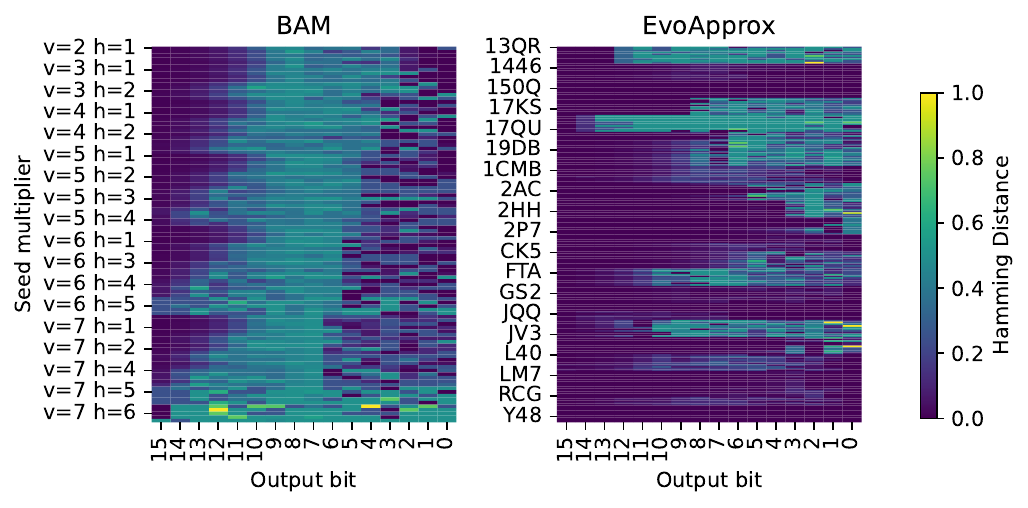}%
    \vspace*{-0.5em}%
    \caption{Average Hamming distance between output bits (0 -- 15) of obfuscated multipliers and their parental multipliers for two classes of circuits: BAM multipliers with different $v$ and $h$ (left) and selected multipliers from EvoApproxLib.}
    \vspace{-1em}
    \label{fig:hamming}
\end{figure}

To get some insight into typical results of the proposed obfuscation algorithm for a given class of multipliers, we analyzed the average Hamming distance between output bits (0 -- 15, one by one) of obfuscated approximate multipliers and their parental approximate multipliers for two classes of circuits: 18 BAM multipliers with different $v$ and $h$ and 19 selected multipliers from EvoApproxLib. 
In the case of BAM multipliers, Fig.~\ref{fig:hamming} shows that logic circuits that compute LSBs are generally not affected much by obfuscations. 
However, the approximate multipliers taken from EvoApproxLib are mostly obfuscated in the LSBs because introducing modifications to MSBs almost always unacceptably degrades the error.

We also investigated the extent to which the obfuscations generated by CGP are unique. Consider a sequence of multipliers $\alpha \vdash a_1 \vdash a_2 \vdash \cdots \vdash a_n \vdash \beta$ generated in a single run by applying mutations. Across all experiments, only 18.7\% obfuscated multipliers are unique, i.e., conditions of uniqueness $\left(\forall n: H_{a_n} \neq H_\beta\right) \wedge H_\beta \neq H_\alpha$ hold. The percentage of unique obfuscations depends on $\alpha$. For approximate circuits (seeds) from EvoApproxLib, BAM, and ICM, the percentage of candidate obfuscations showing the unique behavior is 56.3\%, 14.9\%, and 0.09\%, respectively. 

\subsection{Similarity analysis}
\label{sec:sim:analyz}

Having established that the proposed method can generate obfuscated multipliers with errors similar to $\alpha$ circuits, we now investigate how circuit functionality differs for circuits showing similar error profiles. Since error heat maps $H$ can be treated as images, we employ the Structural Similarity Index Measure (SSIM) to quantify similarity between heat maps and assess functional differences.

\begin{figure}[t]
    \centering
    \includegraphics[width=0.95\columnwidth]{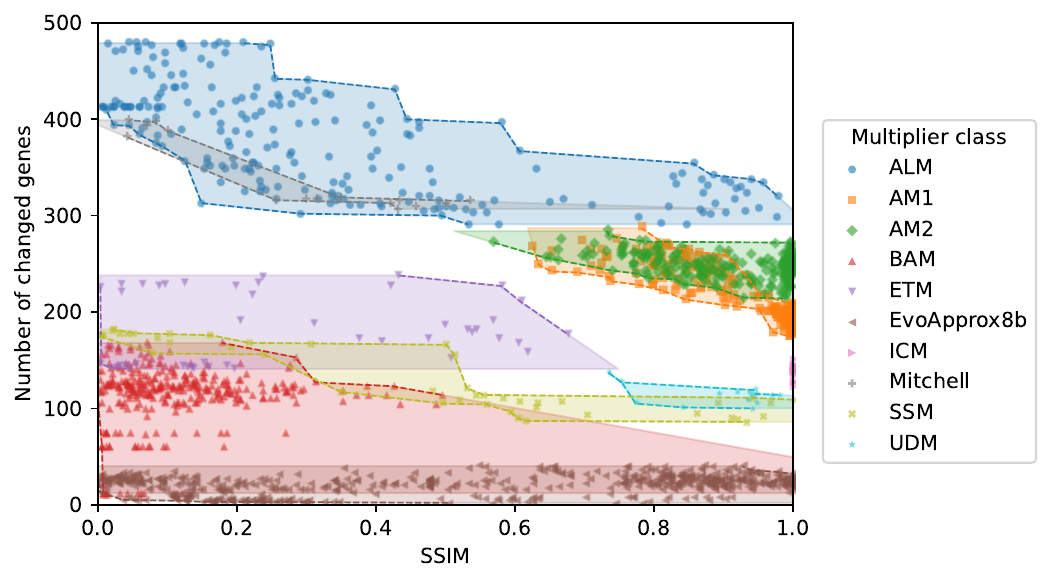}%
    \vspace*{-1.5em}%
    \caption{Properties of generated obfuscated multipliers created from different seeds (denoted as the multiplier classes).}
    \label{fig:params}\vspace{-1em}
\end{figure}

For the obfuscated multipliers created by the proposed method, we examine the relationship between the number of unique modifications (the changed genes) that are performed before reaching the error constraint and the resulting SSIM measured with respect to $\alpha$. Figure~\ref{fig:params} reveals specific behavioral patterns for each class of obfuscated multipliers created from a specific seed.
For example, ALM multipliers, when used as seed, require substantial chromosome changes to reach significant functional differentiation, but the resulting obfuscations can show very diverse SSIM. Obfuscated BAM multipliers consistently result in a low similarity to the original design in terms of SSIM. In contrast, the approximate multipliers from EvoApproxLib, when used as seeds, can be obfuscated to exhibit diverse SSIM (depending on the particular seed); however, the number of acceptable modifications of the seed is low (less than 30); in other words, highly modified circuits required by an attacker are unreachable because their error violates error constraints. 
Obfuscated AM1/2 multipliers maintain a high similarity to their seeds despite extensive modifications. 

\begin{figure}[b]
   \centering
   \vspace{-1em}
   \includegraphics[width=\linewidth]{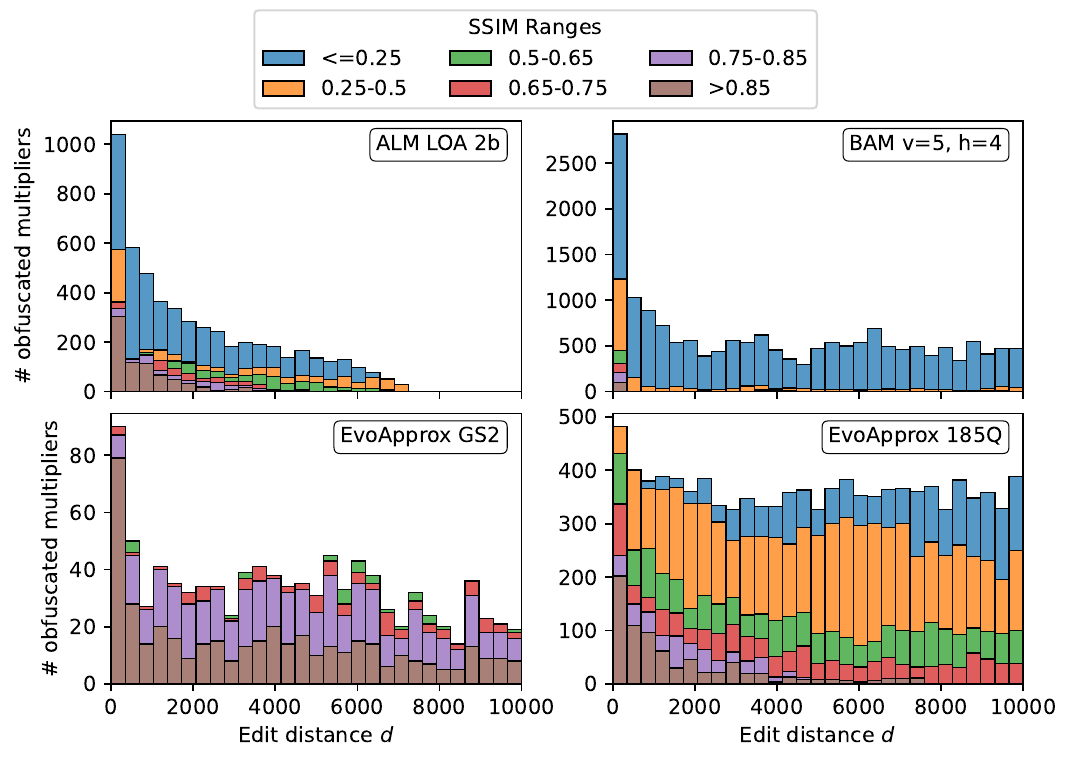}%
   \vspace*{-1.5em}%
   \caption{Distribution of SSIM scores during the obfuscation process (reported in unique steps), where smaller values (blue) indicate high dissimilarity and larger values (brown) represent high visual similarity to the seed multiplier.} 
   \label{fig:dist}
\end{figure}

The distribution of SSIM values during the obfuscation process for investigated multiplier groups is shown in Fig.~\ref{fig:dist}. The ALM multiplier maintains high similarity to the original design for 2k changes ($d$) before transitioning to lower similarity states with SSIM $\leq0.65$. BAM multipliers require only minimal modifications to achieve significant changes in the error heat map. In contrast, obfuscations of the mult8u\_GS2 multiplier show robustness to modifications, consistently maintaining high similarity despite structural changes. 
The absolute count of obfuscated variants indicates that BAM multipliers generate significantly more unique solutions with lower similarity than more irregular designs, such as mult8u\_GS2. It demonstrates that it is more challenging to effectively obfuscate mult8u\_GS2 and other circuits from EvoApproxLib.

\subsection{Classifiers for IP Theft Detection}
\label{sec:class}

This section describes a classification system to determine whether two error heat maps originate from the same source circuit. 

\textit{Setup:} In addition to a simple classification based on one error metric, we utilized Random Forest (RF), Decision Tree (DT), Multilayer-perceptron (MLP), and Support Vector Machine (SVC) classifiers from the scikit-learn library, training them following standard recommended practices.

We also developed a Siamese neural network architecture \cite{SiameseDNN} that processes error heat maps $H_\alpha$ and $H_{\gamma}$ converted to gray-scale images. The network architecture consists of two identical feature extraction branches, each processing 256×256 error heat maps through five convolutional layers. 
The extracted features are then processed through an embedding network with a linear architecture of 32768-512-256-128 neurons, producing 128-dimensional feature embeddings for each input heatmap.
The final classification stage combines the paired embeddings through a classification network with architecture 256-64-32-1, outputting a binary decision score. This architecture contains 18 million trainable parameters. We trained the network using the Adam optimizer with Binary Cross-Entropy loss for 50 epochs.

\textit{Data set construction:} We built two comprehensive data sets to train and evaluate classification models \footnote{The dataset and trained models are publicly available at \url{https://huggingface.co/datasets/ehw-fit/obfax} and \url{https://huggingface.co/ehw-fit/obfax-model}, respectively.}. Each data set comprises pairs of error heat maps $\left(H_{\alpha_1},H_{\beta_1}\right)$  from multipliers $\alpha_1$ and $\beta_1$ with binary labels indicating whether $\alpha_1 \vdash^* \beta_1$ (i.e., whether $\beta_1$ is derived from $\alpha_1$). Each sample is represented as a tensor with dimensions $(N, 2, 256, 256)$, where $N$ denotes the number of pairs of samples. We generate negative samples by pairing each obfuscated circuit with five randomly selected seed circuits $\alpha$ from any circuit family.
The resulting \textit{baseline data set} contains 10,650 samples, partitioned into training (6,750 samples), testing (2,190 samples), and validation (1,710 samples) subsets following standard practices. 

To enhance the robustness of our method, we addressed a potential vulnerability in our approach. Input operand swapping represents one of the simplest obfuscation techniques, where an attacker merely exchanges operands $(a,b)$ to $(b,a)$.
We therefore augmented our data set to include all four possible input permutation variants for both seed and obfuscated multipliers, implemented through diagonal inversion of error heat maps. This expanded \textit{swapped data set} contains $4\times$ the number of samples as the baseline data set.

\begin{table}[b]
    \centering
\caption{Model performance comparison for obfuscation detection on the testing data sets.}
    \label{tab:ml}\vspace{-1em}
\resizebox{\columnwidth}{!}{%
\begin{tabular}{llrrrrrr}
\toprule
\multicolumn{2}{l}{\textbf{Prediction model}}  & \multicolumn{3}{c}{\textbf{Baseline data set}} & \multicolumn{3}{c}{\textbf{Swapped data set}} \\\cline{3-5}\cline{6-8}
    & & \textbf{Acc.} & \textbf{Sens.} & \textbf{Spec.} & \textbf{Acc.} & \textbf{Sens.} & \textbf{Spec.} \\\midrule
\multirow{2}{*}{CL1} & MAE & 62.4\% & 0.2\% & 78.9\% & 62.4\% & 0.2\% & 78.9\% \\
                     & WCE & 63.9\% & 1.1\% & 79.5\% & 63.9\% & 1.1\% & 79.5\% \\\midrule
\multirow{4}{*}{CL2} & DecisionTree & 95.3\% & 92.1\% & 95.9\% & 94.9\% & 73.4\% & 99.2\% \\
                     & MLP & 89.5\% & 49.6\% & 97.5\% & 90.5\% & 71.5\% & 94.2\% \\
                     & RF & 99.6\% & 97.5\% & 100.0\% & 98.2\% & 97.5\% & 98.4\% \\
                     & SVC & 84.2\% & 5.2\% & 100.0\% & 85.0\% & 14.2\% & 99.2\% \\\midrule
\multirow{2}{*}{CL3} & Baseline tr. & 88.9\% & 91.0\% & 88.5\% & 68.4\% & 94.2\% & 63.3\% \\
                     & Swapped tr. & 89.8\% & 96.4\% & 88.5\% & 86.0\% & 96.3\% & 84.0\% \\
\bottomrule
\end{tabular}
}
\end{table}%

\begin{figure}[t]
    \centering
    \includegraphics[width=0.99\columnwidth]{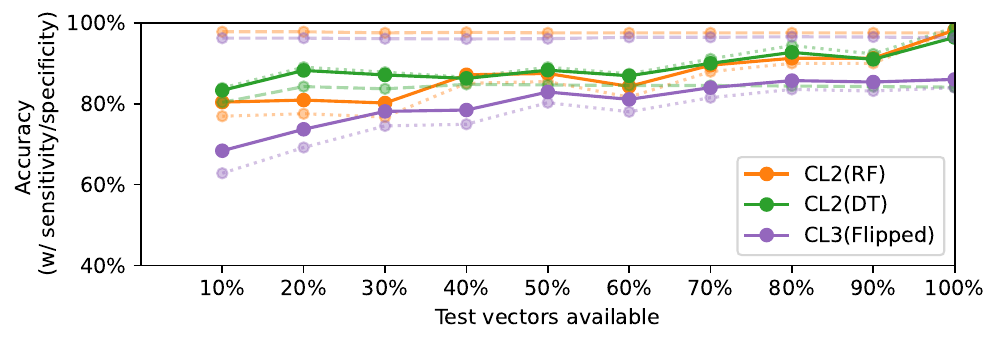}
    \vspace{-1.5em}
  \caption{Detection accuracy when only a subset of vectors is available to calculate the error metrics. The dashed lines indicate sensitivity; the dotted lines represent specificity.}
    \label{fig:subsampled}
    \vspace{-1em}
\end{figure}

\textit{Comparison of proposed classifiers: } Table~\ref{tab:ml} presents the performance evaluation of various detection models on the test data set. The na\"ive approach employing a simple error metric comparison (CL1) demonstrates insufficient sensitivity for reliable IP theft detection. Feature-based models (CL2) achieve excellent performance when implemented with Random Forest classifiers. However, other CL2 models exhibit reduced sensitivity when evaluated on the extended swapped data set. This performance degradation occurs despite the swap-resilient nature of the features used in CL2 classifiers, and importantly, swapped and non-swapped variants are not directly compared within the data set.

The DNN-based classifiers (CL3) demonstrate a relatively high accuracy in detecting similar patterns within error heat maps. However, the composition of the training data set significantly impacts performance generalization. DNN models trained exclusively on the baseline data set exhibit poor specificity when evaluated on the swapped data set, failing to correctly classify input-permuted obfuscations. In contrast, DNN models trained on the complete swapped data set maintain robust performance across the baseline and swapped test scenarios. Although the Random Forest classifier achieves higher overall accuracy than the DNN, the DNN-based method offers distinct advantages for practical deployment, especially for better generalization to previously unseen circuit types and obfuscation strategies not represented in the training data.

\textit{Scalability Issues:} In previous experiments, the circuit's error was obtained by analyzing responses for all possible input combinations (65\,536). In practice, only a subset of responses might be available.  Fig.~\ref{fig:subsampled} shows that the detection accuracy decreases only gradually when the number of test vectors is reduced. For example, the accuracy remains above 80\% for CL2 (decision trees) if only 10\% of the vectors can be used. The results also indicate that CL2 classifiers exhibit greater sensitivity, while the Siamese network (trained on swapped data) shows nearly consistent values if more than 70\% test vectors are available. It has to be noticed that no retraining of the classifiers from Tab.~\ref{tab:ml} was performed for the reduced vector sets.

\section{Conclusions}
\label{sec:conclusions}

We introduced the concept of approximate obfuscation, developed a CGP-based algorithm capable of generating obfuscated circuits, and proposed a framework for reliably detecting obfuscated 8-bit multipliers -- an important class of approximate circuits used in DNN accelerators~\cite{ArmeniakosZSH23}. On our comprehensive dataset, the detection accuracy exceeds 98\%. Future work will focus on improving the method and evaluating it in other use cases.

\begin{acks}
This work was supported by the Czech Science Foundation project no. 26-22525M EvoML-EDA.
\end{acks}

\bibliographystyle{ACM-Reference-Format}
\bibliography{glsvlsi-iptheft}


\begin{thebibliography}{26}


\ifx \showCODEN    \undefined \def \showCODEN     #1{\unskip}     \fi
\ifx \showISBNx    \undefined \def \showISBNx     #1{\unskip}     \fi
\ifx \showISBNxiii \undefined \def \showISBNxiii  #1{\unskip}     \fi
\ifx \showISSN     \undefined \def \showISSN      #1{\unskip}     \fi
\ifx \showLCCN     \undefined \def \showLCCN      #1{\unskip}     \fi
\ifx \shownote     \undefined \def \shownote      #1{#1}          \fi
\ifx \showarticletitle \undefined \def \showarticletitle #1{#1}   \fi
\ifx \showURL      \undefined \def \showURL       {\relax}        \fi
\providecommand\bibfield[2]{#2}
\providecommand\bibinfo[2]{#2}
\providecommand\natexlab[1]{#1}
\providecommand\showeprint[2][]{arXiv:#2}

\bibitem[Ahmed et~al\mbox{.}(2023)]%
        {MAAS:trojan:axc:2023}
\bibfield{author}{\bibinfo{person}{Qazi~Arbab Ahmed}, \bibinfo{person}{Muhammad
  Awais}, {and} \bibinfo{person}{Marco Platzner}.}
  \bibinfo{year}{2023}\natexlab{}.
\newblock \showarticletitle{MAAS: Hiding Trojans in Approximate Circuits}. In
  \bibinfo{booktitle}{\emph{2023 24th Int. Symp. Quality Electronic Design
  (ISQED)}}. \bibinfo{pages}{1--6}.
\newblock
\href{https://doi.org/10.1109/ISQED57927.2023.10129286}{doi:\nolinkurl{10.1109/ISQED57927.2023.10129286}}


\bibitem[Armeniakos et~al\mbox{.}(2023)]%
        {ArmeniakosZSH23}
\bibfield{author}{\bibinfo{person}{Giorgos Armeniakos},
  \bibinfo{person}{Georgios Zervakis}, \bibinfo{person}{Dimitrios Soudris},
  {and} \bibinfo{person}{J{\"{o}}rg Henkel}.} \bibinfo{year}{2023}\natexlab{}.
\newblock \showarticletitle{Hardware Approximate Techniques for Deep Neural
  Network Accelerators: {A} Survey}.
\newblock \bibinfo{journal}{\emph{{ACM} Comput. Surv.}} \bibinfo{volume}{55},
  \bibinfo{number}{4} (\bibinfo{year}{2023}), \bibinfo{pages}{83:1--83:36}.
\newblock
\href{https://doi.org/10.1145/3527156}{doi:\nolinkurl{10.1145/3527156}}


\bibitem[Ceska et~al\mbox{.}(2017)]%
        {ceska:iccad17}
\bibfield{author}{\bibinfo{person}{Milan Ceska}, \bibinfo{person}{Jiri Matyas},
  {et~al\mbox{.}}} \bibinfo{year}{2017}\natexlab{}.
\newblock \showarticletitle{Approximating Complex Arithmetic Circuits with
  Formal Error Guarantees: 32-bit Multipliers Accomplished}. In
  \bibinfo{booktitle}{\emph{Proc. 36th IEEE/ACM Int. Conf. On Computer Aided
  Design}}. \bibinfo{publisher}{IEEE}, \bibinfo{pages}{416--423}.
\newblock
\showISBNx{978-1-5386-3093-8}


\bibitem[Fu et~al\mbox{.}(2024)]%
        {Wille:cgp:2024}
\bibfield{author}{\bibinfo{person}{Rongliang Fu}, \bibinfo{person}{Robert
  Wille}, {and} \bibinfo{person}{Tsung-Yi Ho}.}
  \bibinfo{year}{2024}\natexlab{}.
\newblock \showarticletitle{RCGP: An Automatic Synthesis Framework for
  Reversible Quantum-Flux-Parametron Logic Circuits based on Efficient
  Cartesian Genetic Programming}. In \bibinfo{booktitle}{\emph{Proc. 61st
  ACM/IEEE Design Automation Conf.}} \emph{(\bibinfo{series}{DAC '24})}.
  \bibinfo{publisher}{ACM}.
\newblock
\showISBNx{9798400706011}
\href{https://doi.org/10.1145/3649329.3655950}{doi:\nolinkurl{10.1145/3649329.3655950}}


\bibitem[Jiang et~al\mbox{.}(2019)]%
        {JiangLL019:bookch}
\bibfield{author}{\bibinfo{person}{Honglan Jiang}, \bibinfo{person}{Leibo Liu},
  \bibinfo{person}{Fabrizio Lombardi}, {and} \bibinfo{person}{Jie Han}.}
  \bibinfo{year}{2019}\natexlab{}.
\newblock \showarticletitle{Approximate Arithmetic Circuits: Design and
  Evaluation}.
\newblock In \bibinfo{booktitle}{\emph{Approximate Circuits, Methodologies and
  {CAD}}}. \bibinfo{publisher}{Springer}, \bibinfo{pages}{67--98}.
\newblock


\bibitem[Jiang et~al\mbox{.}(2020)]%
        {Jiang:axc:surv:2020}
\bibfield{author}{\bibinfo{person}{Honglan Jiang}, \bibinfo{person}{Francisco
  Javier~Hernandez Santiago}, \bibinfo{person}{Hai Mo}, \bibinfo{person}{Leibo
  Liu}, {and} \bibinfo{person}{Jie Han}.} \bibinfo{year}{2020}\natexlab{}.
\newblock \showarticletitle{Approximate Arithmetic Circuits: {A} Survey,
  Characterization, and Recent Applications}.
\newblock \bibinfo{journal}{\emph{Proc. {IEEE}}} \bibinfo{volume}{108},
  \bibinfo{number}{12} (\bibinfo{year}{2020}), \bibinfo{pages}{2108--2135}.
\newblock


\bibitem[Liu et~al\mbox{.}(2020)]%
        {Liu:2020}
\bibfield{author}{\bibinfo{person}{Weiqiang Liu}, \bibinfo{person}{Chongyan
  Gu}, \bibinfo{person}{Máire O’Neill}, \bibinfo{person}{Gang Qu},
  \bibinfo{person}{Paolo Montuschi}, {and} \bibinfo{person}{Fabrizio
  Lombardi}.} \bibinfo{year}{2020}\natexlab{}.
\newblock \showarticletitle{Security in Approximate Computing and Approximate
  Computing for Security: Challenges and Opportunities}.
\newblock \bibinfo{journal}{\emph{Proc. IEEE}} \bibinfo{volume}{108},
  \bibinfo{number}{12} (\bibinfo{year}{2020}), \bibinfo{pages}{2214--2231}.
\newblock
\href{https://doi.org/10.1109/JPROC.2020.3030121}{doi:\nolinkurl{10.1109/JPROC.2020.3030121}}


\bibitem[Liu et~al\mbox{.}(2017)]%
        {ALM}
\bibfield{author}{\bibinfo{person}{Weiqiang Liu}, \bibinfo{person}{Jiahua Xu},
  {et~al\mbox{.}}} \bibinfo{year}{2017}\natexlab{}.
\newblock \showarticletitle{Design of Approximate Logarithmic Multipliers}. In
  \bibinfo{booktitle}{\emph{Proc. Great Lakes Symposium on VLSI}}
  \emph{(\bibinfo{series}{GLSVLSI '17})}. \bibinfo{publisher}{ACM},
  \bibinfo{pages}{47–52}.
\newblock
\showISBNx{9781450349727}
\href{https://doi.org/10.1145/3060403.3060409}{doi:\nolinkurl{10.1145/3060403.3060409}}


\bibitem[Mahdiani et~al\mbox{.}(2010)]%
        {Mahdiani:TCSI2009}
\bibfield{author}{\bibinfo{person}{H.~R. Mahdiani}, \bibinfo{person}{A.
  Ahmadi}, {et~al\mbox{.}}} \bibinfo{year}{2010}\natexlab{}.
\newblock \showarticletitle{Bio-Inspired Imprecise Computational Blocks for
  Efficient VLSI Implementation of Soft-Computing Applications}.
\newblock \bibinfo{journal}{\emph{IEEE Trans. Circ. Systems I: Regular Papers}}
  \bibinfo{volume}{57}, \bibinfo{number}{4} (\bibinfo{year}{2010}),
  \bibinfo{pages}{850--862}.
\newblock


\bibitem[Miller(2011)]%
        {miller:cgp:book}
\bibfield{author}{\bibinfo{person}{Julian~F. Miller}.}
  \bibinfo{year}{2011}\natexlab{}.
\newblock \bibinfo{booktitle}{\emph{Cartesian Genetic Programming}}.
\newblock \bibinfo{publisher}{Springer-Verlag}.
\newblock


\bibitem[Mrazek et~al\mbox{.}(2017)]%
        {Mrazek:evoapproxlib:2017}
\bibfield{author}{\bibinfo{person}{Vojtech Mrazek}, \bibinfo{person}{Radek
  Hrbacek}, {et~al\mbox{.}}} \bibinfo{year}{2017}\natexlab{}.
\newblock \showarticletitle{EvoApprox8b: Library of Approximate Adders and
  Multipliers for Circuit Design and Benchmarking of Approximation Methods}. In
  \bibinfo{booktitle}{\emph{Design, Aut. \& Test in Europe Conf. (DATE)}}.
\newblock


\bibitem[Regazzoni et~al\mbox{.}(2018)]%
        {Regazzoni:iccad2018}
\bibfield{author}{\bibinfo{person}{Francesco Regazzoni},
  \bibinfo{person}{Cesare Alippi}, {and} \bibinfo{person}{Ilia Polian}.}
  \bibinfo{year}{2018}\natexlab{}.
\newblock \showarticletitle{Security: The Dark Side of Approximate Computing?}.
  In \bibinfo{booktitle}{\emph{2018 IEEE/ACM Int. Conf. Computer-Aided Design
  (ICCAD)}}. \bibinfo{pages}{1--6}.
\newblock
\href{https://doi.org/10.1145/3240765.3243497}{doi:\nolinkurl{10.1145/3240765.3243497}}


\bibitem[Rostami et~al\mbox{.}(2014)]%
        {Rostami:2014}
\bibfield{author}{\bibinfo{person}{Masoud Rostami}, \bibinfo{person}{Farinaz
  Koushanfar}, {and} \bibinfo{person}{Ramesh Karri}.}
  \bibinfo{year}{2014}\natexlab{}.
\newblock \showarticletitle{A Primer on Hardware Security: Models, Methods, and
  Metrics}.
\newblock \bibinfo{journal}{\emph{Proc. IEEE}} \bibinfo{volume}{102},
  \bibinfo{number}{8} (\bibinfo{year}{2014}), \bibinfo{pages}{1283--1295}.
\newblock
\href{https://doi.org/10.1109/JPROC.2014.2335155}{doi:\nolinkurl{10.1109/JPROC.2014.2335155}}


\bibitem[Scarabottolo et~al\mbox{.}(2020)]%
        {Scarabottolo:pieee:2020}
\bibfield{author}{\bibinfo{person}{Ilaria Scarabottolo},
  \bibinfo{person}{Giovanni Ansaloni}, {et~al\mbox{.}}}
  \bibinfo{year}{2020}\natexlab{}.
\newblock \showarticletitle{Approximate Logic Synthesis: {A} Survey}.
\newblock \bibinfo{journal}{\emph{Proc. IEEE}} \bibinfo{volume}{108},
  \bibinfo{number}{12} (\bibinfo{year}{2020}), \bibinfo{pages}{2195--2213}.
\newblock
\href{https://doi.org/10.1109/JPROC.2020.3014430}{doi:\nolinkurl{10.1109/JPROC.2020.3014430}}


\bibitem[Shafique et~al\mbox{.}(2016)]%
        {Shafique:dac16}
\bibfield{author}{\bibinfo{person}{M. Shafique}, \bibinfo{person}{R. Hafiz},
  {et~al\mbox{.}}} \bibinfo{year}{2016}\natexlab{}.
\newblock \showarticletitle{Invited: Cross-layer approximate computing: From
  logic to architectures}. In \bibinfo{booktitle}{\emph{Proc. of DAC'16}}.
  \bibinfo{pages}{1--6}.
\newblock


\bibitem[Stanley-Marbell et~al\mbox{.}(2020)]%
        {ACsurvey:ACM:2020}
\bibfield{author}{\bibinfo{person}{Phillip Stanley-Marbell},
  \bibinfo{person}{Armin Alaghi}, {et~al\mbox{.}}}
  \bibinfo{year}{2020}\natexlab{}.
\newblock \showarticletitle{Exploiting Errors for Efficiency: {A} Survey from
  Circuits to Applications}.
\newblock \bibinfo{journal}{\emph{ACM Comput. Surv.}} \bibinfo{volume}{53},
  \bibinfo{number}{3} (\bibinfo{year}{2020}), \bibinfo{numpages}{39}~pages.
\newblock
\showISSN{0360-0300}


\bibitem[Taigman et~al\mbox{.}(2014)]%
        {SiameseDNN}
\bibfield{author}{\bibinfo{person}{Yaniv Taigman}, \bibinfo{person}{Ming Yang},
  \bibinfo{person}{Marc'Aurelio Ranzato}, {and} \bibinfo{person}{Lior Wolf}.}
  \bibinfo{year}{2014}\natexlab{}.
\newblock \showarticletitle{DeepFace: Closing the Gap to Human-Level
  Performance in Face Verification}. In \bibinfo{booktitle}{\emph{2014 IEEE
  Conf. Comp. Vision and Pattern Recognition}}. \bibinfo{pages}{1701--1708}.
\newblock


\bibitem[Ullah et~al\mbox{.}(2018)]%
        {SMApproxlib:dac18}
\bibfield{author}{\bibinfo{person}{Salim Ullah},
  \bibinfo{person}{Sanjeev~Sripadraj Murthy}, {and} \bibinfo{person}{Akash
  Kumar}.} \bibinfo{year}{2018}\natexlab{}.
\newblock \showarticletitle{{SMApproxlib}: Library of {FPGA}-Based Approximate
  Multipliers}. In \bibinfo{booktitle}{\emph{Proc. 55th Annual Design
  Automation Conference}}. \bibinfo{publisher}{ACM}, \bibinfo{address}{New
  York, NY, USA}, Article \bibinfo{articleno}{157},
  \bibinfo{numpages}{6}~pages.
\newblock


\bibitem[Vasicek(2019)]%
        {vasicek:access2019}
\bibfield{author}{\bibinfo{person}{Zdenek Vasicek}.}
  \bibinfo{year}{2019}\natexlab{}.
\newblock \showarticletitle{Formal Methods for Exact Analysis of Approximate
  Circuits}.
\newblock \bibinfo{journal}{\emph{IEEE Access}} \bibinfo{volume}{7},
  \bibinfo{number}{1} (\bibinfo{year}{2019}), \bibinfo{pages}{177309--177331}.
\newblock


\bibitem[Vijayakumar et~al\mbox{.}(2017)]%
        {Sandip:2017}
\bibfield{author}{\bibinfo{person}{Arunkumar Vijayakumar},
  \bibinfo{person}{Vinay~C. Patil}, {et~al\mbox{.}}}
  \bibinfo{year}{2017}\natexlab{}.
\newblock \showarticletitle{Physical Design Obfuscation of Hardware: A
  Comprehensive Investigation of Device and Logic-Level Techniques}.
\newblock \bibinfo{journal}{\emph{IEEE Trans. Inf. Forensics Security}}
  \bibinfo{volume}{12}, \bibinfo{number}{1} (\bibinfo{year}{2017}),
  \bibinfo{pages}{64--77}.
\newblock


\bibitem[Witschen et~al\mbox{.}(2019)]%
        {WitschenMAP19}
\bibfield{author}{\bibinfo{person}{Linus Witschen}, \bibinfo{person}{Hassan~G.
  Mohammadi}, {et~al\mbox{.}}} \bibinfo{year}{2019}\natexlab{}.
\newblock \showarticletitle{Jump Search: {A} Fast Technique for the Synthesis
  of Approximate Circuits}. In \bibinfo{booktitle}{\emph{Proc. 2019 Great Lakes
  Symposium on VLSI, {GLSVLSI}}}. \bibinfo{publisher}{{ACM}},
  \bibinfo{pages}{153--158}.
\newblock


\bibitem[Wu et~al\mbox{.}(2024)]%
        {Wu:axmult:acm:surv:2024}
\bibfield{author}{\bibinfo{person}{Ying Wu}, \bibinfo{person}{Chuangtao Chen},
  {et~al\mbox{.}}} \bibinfo{year}{2024}\natexlab{}.
\newblock \showarticletitle{A Survey on Approximate Multiplier Designs for
  Energy Efficiency: From Algorithms to Circuits}.
\newblock \bibinfo{journal}{\emph{ACM Trans. Des. Autom. Electron. Syst.}}
  \bibinfo{volume}{29}, \bibinfo{number}{1} (\bibinfo{date}{Jan.}
  \bibinfo{year}{2024}), \bibinfo{numpages}{37}~pages.
\newblock
\showISSN{1084-4309}
\href{https://doi.org/10.1145/3610291}{doi:\nolinkurl{10.1145/3610291}}


\bibitem[Xie and Srivastava(2019)]%
        {antiSAT:attack:2019}
\bibfield{author}{\bibinfo{person}{Yang Xie} {and} \bibinfo{person}{Ankur
  Srivastava}.} \bibinfo{year}{2019}\natexlab{}.
\newblock \showarticletitle{Anti-SAT: Mitigating SAT Attack on Logic Locking}.
\newblock \bibinfo{journal}{\emph{IEEE Trans. Comput.-Aided Design Integr.
  Circuits Syst.}} \bibinfo{volume}{38}, \bibinfo{number}{2}
  (\bibinfo{year}{2019}), \bibinfo{pages}{199--207}.
\newblock
\href{https://doi.org/10.1109/TCAD.2018.2801220}{doi:\nolinkurl{10.1109/TCAD.2018.2801220}}


\bibitem[Yasaei et~al\mbox{.}(2021)]%
        {GNN4IP:2021}
\bibfield{author}{\bibinfo{person}{Rozhin Yasaei}, \bibinfo{person}{Shih-Yuan
  Yu}, {et~al\mbox{.}}} \bibinfo{year}{2021}\natexlab{}.
\newblock \showarticletitle{GNN4IP: Graph Neural Network for Hardware
  Intellectual Property Piracy Detection}. In \bibinfo{booktitle}{\emph{58th
  ACM/IEEE Design Automation Conference (DAC)}}. \bibinfo{pages}{217--222}.
\newblock
\href{https://doi.org/10.1109/DAC18074.2021.9586150}{doi:\nolinkurl{10.1109/DAC18074.2021.9586150}}


\bibitem[Yellu and Yu(2023)]%
        {Yellu:2023}
\bibfield{author}{\bibinfo{person}{Pruthvy Yellu} {and}
  \bibinfo{person}{Qiaoyan Yu}.} \bibinfo{year}{2023}\natexlab{}.
\newblock \showarticletitle{Securing Approximate Computing Systems via
  Obfuscating Approximate-Precise Boundary}.
\newblock \bibinfo{journal}{\emph{IEEE Trans. Comput.-Aided Design Integr.
  Circuits Syst.}} \bibinfo{volume}{42}, \bibinfo{number}{1}
  (\bibinfo{year}{2023}), \bibinfo{pages}{27--40}.
\newblock
\href{https://doi.org/10.1109/TCAD.2022.3168261}{doi:\nolinkurl{10.1109/TCAD.2022.3168261}}


\bibitem[Yi et~al\mbox{.}(2025)]%
        {Yi:GPTAC:2024}
\bibfield{author}{\bibinfo{person}{Sipei Yi}, \bibinfo{person}{Weichuan Zuo},
  {et~al\mbox{.}}} \bibinfo{year}{2025}\natexlab{}.
\newblock \showarticletitle{GPTAC: Domain-Specific Generative Pre-Trained Model
  for Approximate Circuit Design Exploration}.
\newblock \bibinfo{journal}{\emph{IEEE J. Emerging Selected Topics in Circ. and
  Systems}} \bibinfo{volume}{15}, \bibinfo{number}{2} (\bibinfo{year}{2025}),
  \bibinfo{pages}{349--360}.
\newblock


\end{thebibliography}

\end{document}